\documentclass[twocolumn,nofootinbib,amsmath,amssymb,aps,prl,balancelastpage,superscriptaddress,floatfix]{revtex4-1}
\usepackage{lipsum}

\usepackage{graphicx}

\usepackage{dcolumn}
\usepackage{bm}

\usepackage[normalem]{ulem}

\usepackage{blindtext}
\usepackage[T1]{fontenc}
\usepackage{verbatim}
\usepackage{letltxmacro}
\newcommand{\cmt}[1]{}

\newcommand\wordcount{\verbatiminput{\jobname.sum}}

\newif\ifcorrectingmode
\correctingmodetrue

\LetLtxMacro\origcite\cite

\renewcommand{\cite}[2][]{%
  \ifcorrectingmode
  \mbox{\origcite[#1]{#2}}%
  \else
  \origcite[#1]{#2}%
  \fi
}
\usepackage{hyperref}
\usepackage{slashed}
\usepackage{amsfonts}
\usepackage{mathrsfs}
\usepackage{color}
\usepackage[makeroom]{cancel}
\usepackage{graphicx}
\usepackage{amssymb}
\usepackage{amsmath}
\usepackage{hyperref}
\usepackage{cleveref}

\def\p{\partial}

\def\bi{\begin{itemize}}
\def\ei{\end{itemize}}
\def\be{\begin{equation}}
\def\ee{\end{equation}}
\newcommand{\bea}{\begin{eqnarray}}
\newcommand{\eea}{\end{eqnarray}}

\begin{document}

\begin{flushright} {\footnotesize  YITP-25-100}  \end{flushright}


\title{Hybrid Inflation from Fermion Condensation}
\author{Stephon~Alexander}
\affiliation{Brown Theoretical Physics Center and Department of Physics,Brown University,RI 02903, USA}

\author{Pisin~Chen}
\affiliation{Leung Center for Cosmology and Particle Astrophysics, National Taiwan University, Taipei 10617, Taiwan}
\affiliation{Department of Physics and Graduate Institute of Astrophysics, National Taiwan University, Taipei 10617, Taiwan}
\affiliation{Kavli Institute for Particle Astrophysics and Cosmology, SLAC National Accelerator Laboratory, Stanford University, CA 94305, U.S.A.}

\author{Jinglong~Liu}
\affiliation{Center for Field Theory and Particle Physics \& Department of Physics, Fudan University, 200433 Shanghai, China}
\affiliation{Center for Astronomy and Astrophysics, Fudan University, 200433 Shanghai, China}
\affiliation{Tsung-Dao Lee Institute, 1 Lisuo Road, Shanghai, 201210, China}
\affiliation{School of Physics and Astronomy, Shanghai Jiao Tong University, 800 Dongchuan Road, Shanghai 200240, China}

\author{Antonino~Marcian\`o}
\affiliation{Center for Field Theory and Particle Physics \& Department of Physics, Fudan University, 200433 Shanghai, China}
\affiliation{Center for Astronomy and Astrophysics, Fudan University, 200433 Shanghai, China}
\affiliation{Laboratori Nazionali di Frascati INFN, Frascati (Rome), Italy, EU}
\affiliation{INFN sezione Roma Tor Vergata, I-00133 Rome, Italy, EU}

\author{Misao~Sasaki}
\affiliation{Kavli Institute for the Physics and Mathematics of the Universe (WPI), The University of Tokyo Institutes for Advanced Study, The University of Tokyo, Chiba 277-8583, Japan}
\affiliation{Asia Pacific Center for Theoretical Physics, Pohang 37673, Korea}
\affiliation{Center for Gravitational Physics, Yukawa Institute for Theoretical Physics, Kyoto University, Kyoto 606-8502, Japan}
\affiliation{Leung Center for Cosmology and Particle Astrophysics, National Taiwan University, Taipei 10617, Taiwan}

\author{Xuan-Lin~Su}
\affiliation{Center for Field Theory and Particle Physics \& Department of Physics, Fudan University, 200433 Shanghai, China}
\affiliation{Center for Astronomy and Astrophysics, Fudan University, 200433 Shanghai, China}

\begin{abstract}
\noindent 
We investigate how inflation can emerge from four-fermion interactions generated by spacetime torsion, eliminating the need for additional scalar fields beyond the Standard Model. We partition fermions in two sectors and introduce two bound fields. In the effective theory approach, once all the fermions have been integrated out, the bound fields serve as the inflaton and the auxiliary field, in analogy to the hybrid inflation and accounting for a waterfall (hybrid) mechanism. The inclusion of an axial chemical potential naturally facilitates the end of reheating. During the waterfall regime, the effective potential governing the fermion condensate supports the formation of non-topological solitons, known as Q-balls, which can be accounted for seeding primordial black holes (PBHs). 
\end{abstract}

\maketitle

\emph{Introduction.} --- The inflationary paradigm \cite{Guth:1980zm,STAROBINSKY198099} posits an early phase of accelerated expansion driven by a scalar field, resolving key cosmological issues and generating primordial density fluctuations \cite{LINDE1983177,Albrecht:1982wi}. Hybrid inflation, introduced by Linde \cite{Linde:1993cn}, combines chaotic inflation \cite{LINDE1983177} with symmetry-breaking potentials. It involves a slowly rolling inflaton field and a secondary field, which undergoes a rapid ``waterfall'' transition once the inflaton crosses a critical threshold, terminating inflation. This process can lead to primordial black hole (PBH) formation \cite{
%
Chen:2002tu,Chen:2004ft}, with PBHs proposed as dark matter candidates and early universe probes \cite{Zeldovich:1967lct,hawking1971gravitationally,carr1974black,1975ApJ...201....1C,Zeldovich:1967lct,Garcia-Bellido:1996mdl,Khlopov:2008qy,Frampton:2010sw,Kawasaki:2016pql,Carr:2016drx,Inomata:2016rbd,Inomata:2017okj,Georg:2017mqk,Chen:2002tu,Chen:2004ft,Adler:2001vs}. 
PBHs may also originate from Q balls \cite{Cotner:2019ykd,Cotner:2016dhw,Cotner:2017tir}, stable scalar configurations with conserved charge that emerge in theories with attractive self-interactions. They can influence reheating and structure formation \cite{Enqvist:1998en,Kasuya:2001hg,Kasuya:2014ofa} and have been studied as dark matter candidates in supersymmetric models \cite{Kusenko:1997si,Kusenko:1997zq,Enqvist:1998xd}.

Many models of inflation rooted in fundamental scalar fields encounter a persistent challenge: ultraviolet divergences as one approaches trans-Planckian scales at the onset of inflation. By contrast, in a composite framework such as fermion condensation, this issue is naturally alleviated --- once the scale of compositeness is reached, the effective description ceases to be valid, preventing unphysical extrapolations. Motivated by this, we delve into fermion-condensate inflation and show how the model not only circumvents these UV difficulties but also naturally incorporates dark matter, all without invoking any scalar degrees of freedom beyond the Standard Model. In this picture, fermions themselves play a central role in the early universe, undergoing Bardeen-Cooper-Schrieffer (BCS) condensation under attractive interactions \cite{Alexander:2009uu,Shapiro:2001rz,Tong:2023krn}, which can be described by the Nambu--Jona-Lasinio model \cite{Inagaki:1993ya}. This generates an effective Coleman-Weinberg-like potential that can drive inflation \cite{Addazi:2017qus}.

Here we extend the model of fermion condensate inflation \cite{Addazi:2017qus} to a hybrid-inflation-like model by considering two sectors of the fermions. These two sectors will introduce two bound fields, separately corresponding to the inflaton and the auxiliary field in hybrid inflation model, the latter of which undergoes the waterfall mechanism and naturally explains the formation of PBH without postulating the existence of new scalar fields \cite{Garcia-Bellido:1996mdl,Chen:2004ft}.\\

\emph{Hybrid inflation from fermion condensate.} --- 
{Moving along the lines of Ref.~\cite{Addazi:2017qus}, we consider two separate fermion sectors: the set of fermion fields $\{\Psi_i\}$, with $i=1\dots N$, denoted as $\psi$; and the set of fermion fields $\{\Psi_j\}$, with $j=1\dots \bar{N}$, denoted as $\chi$. The action for each sector is recovered by summing the contribution due to each constituent fermionic field, i.e. 
\begin{equation}
 \mathcal{S}[\psi]=\sum_{i=1}^{N}\mathcal{S}[{\Psi_i}] 
 \,,
\qquad 
\mathcal{S}[\chi]=\sum_{j=1}^{\bar{N}}\mathcal{S}[{\Psi_j}]\,,    
\end{equation}
where each fermion field $\Psi_i$ is governed by the Dirac action
\begin{equation}
    \mathcal{L}_{\rm Dirac} [\Psi_i]= \bar{\Psi}_i\gamma^I e^\mu_I\left(1-\frac{\imath}{\beta}\gamma^5\right)\imath \nabla_\mu\Psi_i  +{\rm  h.c.}\,,
\end{equation}
with non-minimal coupling term inverse proportional to} 
{$\beta\in \mathbb{R}$}, 
{covariant derivative given by $\nabla_\mu = \p_\mu + \frac{1}{4}\omega^{IJ}_\mu[\gamma_I,\gamma_J]$, and where the $i$ indices are not summed over.} \\

{We can recast the total Dirac Lagrangian $\mathcal{L}_{\rm Dirac}'[\psi, \chi]$ by integrating out the contorsion tensor \cite{PhysRevD.73.044013,Alexander:2014eva,Alexander:2014uaa,1971PhLA...36..225H,hehl1973spin,hehl1974spin,lord1976tensors,doi:10.1142/0233,de1994spin,Shapiro:2001rz,hammond2002torsion,Bambi:2014uua}, namely, 
\begin{align}
    &\mathcal{L}_{\rm Dirac}'[\psi, \chi] = \bar{\psi}\gamma^I e^\mu_I \imath \tilde{\nabla}_\mu\psi+ \bar{\chi}\gamma^I e^\mu_I \imath \tilde{\nabla}_\mu\chi+{\rm h.c.} \,,\label{eq.LDirac}\\
    &\mathcal{L}_{\rm int} [\psi, \chi] = - \xi\kappa^2 J_5^LJ_5^M\eta_{LM},\label{eq.Lint}
\end{align}
where $J^I_5\!=\!\bar{\psi}\gamma_5 \gamma^I \psi + \bar{\psi}\gamma_5 \gamma^I \chi+  \bar{\chi}\gamma_5 \gamma^I \psi+ \bar{\chi}\gamma_5 \gamma^I \chi$, $\tilde{\nabla}_\mu$ denotes the torsion-free and metric-compatible covariant derivative,} 
{the dimensional coupling constant $\kappa^2=8 \pi G$ has been introduced as a function of the Newton constant $G$,}  
{and $\xi$ is a constant factor that is a combination of $\beta$ and $\gamma$,} 
{the latter being the Barbero-Immirzi parameter that appears in the (topological) gravitational Holst term.} 
{The value of $\xi$ has been discussed in Ref.~\cite{Alexander:2014eva}: the prefactor of the interaction term $-\xi\kappa^2$ may provide a strong and attractive self-interaction for fermions and hence fermion condensation. The Fierz identity and the Pauli-Kofink relation \cite{PhysRev.140.B1467,10.1007/978-94-011-1719-7_6} can be used to simplify the axial current square interaction into
\begin{eqnarray} \label{gregorius}
 &&J^L_5 J^M_5  \eta_{LM} =  
 2(\bar{\psi}\psi)^2 - 2(\bar{\psi}\gamma_5 \psi)^2 +  2(\bar{\chi}\chi)^2 - 2(\bar{\chi}\gamma_5\chi)^2 
 \nonumber \\
&& + 2 \,\bar{\psi} \chi \,\bar{\chi}\psi 
+ 2\bar{\psi} \psi \,\bar{\chi}\chi - 2\bar{\psi} \gamma_5 \psi \, \bar{\chi} \gamma_5  \chi - 2 \,\bar{\psi} \gamma_5  \chi \,\bar{\chi} \gamma_5 \psi \,.
\end{eqnarray}
}

{The quantum theory is hence defined by the path integral 
\begin{equation}\label{patpsichi}
 \mathcal{Z}[\psi, \chi] = \int \mathcal{D}\psi \mathcal{D}\chi \, e^{\imath (\mathcal{S}'_{\rm Dirac}[\psi, \chi] + \mathcal{S}_{\rm int}[\psi,\chi]) }
 \,,   
\end{equation}
with $\mathcal{S}'_{\rm Dirac}[\psi, \chi]=\mathcal{S}[\psi]+\mathcal{S}[\chi]$. Following Ref.~\cite{Alexander:2022cow}, we can multiply the path-integral Eq.~\eqref{patpsichi} by a constant term}
\begin{align}
    \mathcal{Z}_\alpha = \int \mathcal{D}\alpha \mathcal{D}\bar{\alpha} \exp\left(-\int d^4x M_\phi^{-2}\bar{\alpha}\alpha\right)\,,
\end{align}
where $\alpha$ is an auxiliary field and $M_\phi$ is a mass term, corresponding to a bound scalar field defined by $\phi = \alpha - M_\phi^{-2} \bar{\chi}_{R}\chi_{L} \equiv \sigma + \imath \varpi$. Then, after integrating out the $\chi$ fermions at one-loop level according to the calculations provided by Ref.~\cite{Alexander:2022cow}, the effective action becomes
\begin{align}\label{Lefftwof}
    &\tilde{\mathcal{L}}_\psi = \bar{\psi}(\imath \gamma^\mu\nabla_\mu) \psi + \frac{1}{2 m^2}[(\bar{\psi}\psi)^2 + (\imath \bar{\psi}\gamma^5\psi)^2] \\
    & + (\p_\mu\phi^*)(\p^\mu\phi) +  y(\phi\bar{\psi}_L\psi_R + \textrm{h.c.}) - m_\phi^2|\phi|^2 - \frac{\lambda_\phi}{4}|\phi|^4,\notag
\end{align}
where 
{we have defined $1/(2m^2) = -2\xi\kappa^2$,} and the effective parameters $y$, $m_\phi$, and $\lambda_\phi$ arise from integrating out $\chi$ at one-loop level up to the energy cut-off $\Lambda$ --- the result was provided in Ref.~\cite{Alexander:2022cow}. 
We adopt representative values for the effective couplings: $y \approx \sqrt{2}\pi$, $m_\phi^2 \approx \Lambda^2$, 
{while $\lambda_\phi$ is vanishing, due to dimensional regularization.}\\ 

{As an analogy to the Nambu--Jona-Lasinio model, we can introduce in the Dirac action two gap fields for the scalar and pseudoscalar term, respectively, by means of the Hubbard-Stratonovich (HS) transform in the partition function, which is provided by \cite{Tong:2023krn,Alexander:2009uu,Shapiro:2001rz,alexander2009cosmological,Inagaki:1993ya,Addazi:2017qus}
\begin{align}\label{Partition}
\mathcal{Z}_{\Sigma, \Pi}[\psi, \bar{\psi}] = \int \mathcal{D}\Sigma\mathcal{D}\Pi\mathcal{D}\bar{\psi}\mathcal{D}\psi e^{\imath \mathcal{S}_{\Sigma, \Pi}\left[\bar{\psi},\psi \right]},
\end{align}
with
\begin{align}\label{SCon}
&\mathcal{S}_{\Sigma,\Pi}[ \psi, \bar{\psi}] = \int d^4x |e| \left[\frac{1}{2}\bar{\psi}(\gamma^Ie^\mu_I \imath \nabla_\mu)\psi+{\rm h.c.}\right.\notag\\
    &\left.
-\frac{m^2}{2}(\Sigma^2+\Pi^2)-\bar{\psi}(\Sigma+\imath \gamma^5\Pi+y\sigma+y\imath\gamma^5\varpi)\psi\right] \\
    &
    +(\p_\mu\phi^*)(\p^\mu\phi) + m_\phi^2|\phi|^2 
    . \notag
\end{align}
In Eq.~\eqref{SCon} the non-exact axial $U(1)_A$ symmetry $\psi\rightarrow e^{\imath \gamma^5\theta}\psi$  allows us to introduce a parity odd chemical potential, namely the axial chemical potential $\mu$ by $\mathcal{L}_\mu = -\mu\bar{\psi}\gamma^0\gamma^5\psi$, which, e.g., has been discussed in~\cite{Tong:2023krn}. For convenience, we define $\mathbf{A}\equiv\Sigma+\imath \gamma^5\Pi\equiv Ae^{\imath \gamma^5\theta_A}$, where $A^2 = \Sigma^2 + \Pi^2$, and $\theta = \arctan\left({\Pi}/{\Sigma}\right)$. We shift away the exponential term by using $\psi\rightarrow e^{-\imath \gamma^5\theta/2}\psi$.
}\\

\begin{figure}[htbp]
    \centering
    \includegraphics[width=0.45\textwidth]{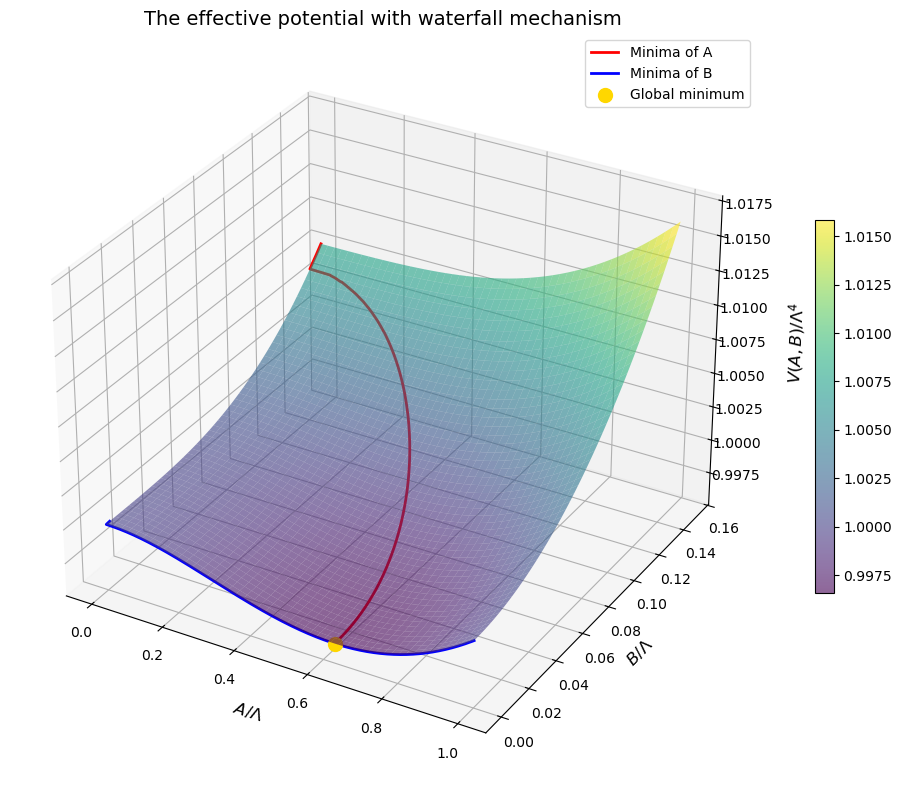}
    \caption{We plot the normalized effective potential $V(A,B)/\Lambda^4$, with the  $B$-axis truncated at $0.2\,\Lambda$ for a clearer view of the minima. The minima of $A$ are highlighted in red, while those of $B$ in blue, the other parameters in the potential being kept constant. With the parameters hence specified, the figure shows that the fermion condensate field $A$ undergoes a phase transition, when the bound field $B$, acting as the inflaton, is rolling approaching zero. The golden color point marks the global minimum of the potential, located at the intersection of the two curves of minima. This observation supports the presence of a waterfall mechanism.}
    \label{fig:Potential}
\end{figure}

Now we introduce a composite field $B \equiv \sigma + i\gamma^5 \varpi \equiv |B| e^{i\gamma^5 \theta_B}$, analogous to the field $A$ used for $\psi$. For simplicity, we consider a configuration where the field $A$ contains only the scalar component, while $B$ includes only the pseudoscalar part. This can be realized by appropriately choosing the phases $\theta_A$ and $\theta_B$. Using $|B| = |\phi|$, the effective potential, {
after integrating out the fermion field $\psi$ following Refs.~\cite{Addazi:2017qus,Inagaki:1993ya},} recasts as 
\begin{align}
  \bar{V}(A,B) = V(A^2+y^2B^2) - m_\phi^2|B|^2 
  ,
\end{align}
where $V(F^2)$ is the effective potential 
{\begin{align}\label{effpotential}
    & V(F^2) = V_0 + \frac{m^2}{2}F^2 - \frac{1}{4\pi^2} \left[(F^2-\mu^2)\Lambda^2 + \right. \\
    & \left. \Lambda^4\ln\left(1+\frac{F^2-\mu^2}{\Lambda^2}\right) - (F^2-\mu^2)^2\ln\left(1+\frac{\Lambda^2}{F^2-\mu^2}\right)\right]\notag\\
    & \!+\! \frac{1}{(4\pi)^2}\frac{R}{6}\left[(F^2-\mu^2)\ln\left(\!1\! +\!\frac{\Lambda^2}{F^2-\mu^2}\right)\!\!-\!\!
    \frac{\Lambda^2(F^2-\mu^2)}{\Lambda^2+F^2-\mu^2}\!\right]\,. \notag
\end{align}
}
We follow Ref.~\cite{Addazi:2017qus} to choose $V_0 = \Lambda^4$ and $\Lambda^2 = 2\pi^2 m^2$, and we inspect the minimum of the normalized potential $V(A,B)/\Lambda^4$ for $\tilde{A}$ and $\tilde{B}$, where tilde indicates the field to be normalized with the energy cutoff $\Lambda$. The zeros are provided, respectively, by $\tilde{B} = 0$ and $\tilde{A} = 0$ or $A^2 \approx 0.398 + \tilde{\mu}^2 - y^2\tilde{B}^2$, where we have ignored the contribution of the terms proportional to {
$R$, being $R=6H^2$ in the FLRW metric and $H^2\ll \Lambda^2$}. We find that the minimal values of $B$ are always located at zero, with their running triggering non-zero solutions for $\tilde{A}$. {
Hence a phase transition happens and the field $A$ starts to fall from its initial zero value to the new minimum, providing a waterfall mechanism.} This is shown in Fig.~\eqref{fig:Potential}. As a result, as the field $B$ decreases, a waterfall mechanism is triggered for the field $A$.
After the end of the slow roll phase, the gap field and the bound field relax to their minima. This implies for the bound field that $B=0$, and that the gap field oscillates around its minimum $A_0 \approx 0.6\,\Lambda$, in correspondence to which the fermion condensate is formed as a mean-field approximation of the gap field, and the condensate satisfies the gap equation $\p V(A)/\p A =0$. 
{Finally, we find the relations} 
\begin{align}
    \langle n_A\rangle &= \langle \bar{\psi}\gamma^0\gamma^5\psi\rangle = \left\langle \frac{\partial U}{\partial \mu} \right \rangle
    \,,\label{prim}\\
    \langle \imath \bar{\psi}\gamma^5\psi\rangle &= \left\langle \frac{\partial U(A)}{\partial \Pi} \right \rangle 
    =  \left\langle \frac{\partial U(A)}{\partial A}\frac{\partial A}{\partial \Pi} 
    \right\rangle
    \,,\label{psconeq}
\end{align}
{where we have defined $U(A)=V(A)-m^2 A^2/2$, and calculated the expectation values of the fermionic bilinears, respectively, in Eq.~\eqref{prim} from the derivative of the generating  functional that involves the chemical potential, and in Eq.~\eqref{psconeq} from the derivative of the partition function, Eq.~\eqref{Partition}, which involves the gap field $\Pi$.}
Eq.~\eqref{prim} provides the equation of motion for the axial chemical potential, while Eq.~\eqref{psconeq} provides the expression for the chiral anomaly, together with the equation of motion of the Noether current for $U(1)_A$, namely
\begin{align}\label{axialcurrenteq}
    \nabla_\mu \langle J^{\mu}_5\rangle = 2A_0\langle \imath \bar{\psi}\gamma_5\psi\rangle\,,
\end{align}
where $J^{\mu}_5=\bar{\psi}\gamma^\mu\gamma_5\psi$ is the axial fermion current 
{and $A_0$, the minimum of the gap field, provides the correction to the mass term appearing in the chiral anomaly}. Then combining Eq.~\eqref{axialcurrenteq} and Eq.~\eqref{psconeq}, we recover the chiral anomaly in the Friedmann-Lema\^itre-Robertson-Walker (FLRW) metric, i.e. 
\begin{align}\label{anomalyeq}
    \p_\tau\langle n_A\rangle + 3\mathcal{H}\langle n_A\rangle = 2 m^2 A_0^2 a \,,
\end{align}
where we have expressed the equation in the conformal time $\tau = \mathcal{H}^{-1} = 1/(aH)$. The solution to Eq.~\eqref{anomalyeq}, using $a \propto \tau$ during the radiation dominated universe, is found to {
include two parts: the first part is proportional to $\tau^{-3}$ and will be diluted quickly, while the second part is proportional to $\tau^2$ and will increase.} Thus, the number density $\langle n_A\rangle$ increases.
\begin{figure}[htbp]
    \centering
    \includegraphics[width=0.4\textwidth]{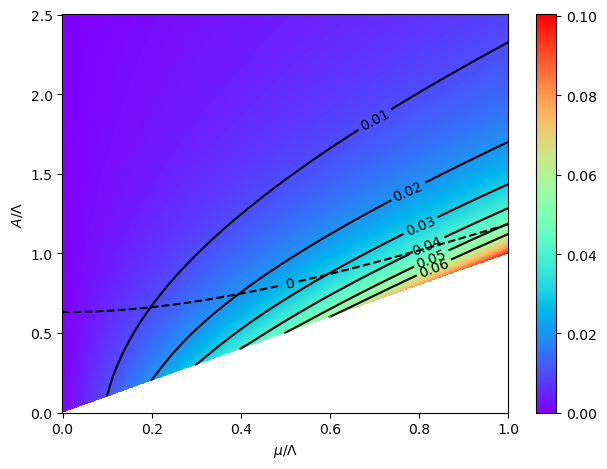}
    \caption{The value of the right-hand side of the equation of motion for the chemical potential is expressed in terms of shades of colors. The solid curves represent specific values of the normalized number density. Solutions to the gap equation lie on the dashed curve. The intersections between the dashed curves and the solid curves are the solutions allowed, with specific number densities.}
 \label{fig:condensateEOM}
\end{figure}
We numerically solve the gap equation $\p V(A)/\p A =0$ and the chemical potential equation, Eq.~\eqref{prim}, due to the increase in number density, and show the result in Fig.~\eqref{fig:condensateEOM}. The colors represent the values on the right hand side of the chemical potential equation, Eq.~\eqref{prim}, with the dashed curve given by the solution of the gap equation. The solid curves represent the solutions with specific values of the number density normalized by the energy cutoff $\langle n_A\rangle / \Lambda^3$. The intersections between the dashed and the solid curves provide the solutions for the fermion condensate. We can see from Fig.~\eqref{fig:condensateEOM} that when $\langle n_A\rangle /\Lambda^3 \gtrsim 0.05$, which is between the two solid curves on the far right, there exist no solutions that enable condensation. This implies that the effective potential in Eq.~\eqref{effpotential} is not viable any longer. This scenario finally provides a mechanism for the evaporation of the condensate, and thus the end of reheating.\\

\emph{Condensate fragmentation.} --- We discuss the structure formation in the waterfall regime, choosing $B=0$. $A$ falling from zero to the true vacuum provides for the gap fields the identification with the fermion condensates at equilibrium, i.e. $\Sigma_0 = \langle h |\bar{\psi}\psi|h\rangle$ and $\Pi_0=\langle h|\bar{\psi}\imath \gamma^5\psi|h\rangle$. We may now define a new field sharing the same radial part of $A$, i.e. $\varphi = \psi^\dagger_R\psi_L/(\sqrt{2}m^2)$ and $\varphi^* = \psi_L^\dagger\psi_R /(\sqrt{2}m^2).$
We then have $\Sigma = \sqrt{2}(\varphi+\varphi^*)$, $\Pi = \sqrt{2}\imath (\varphi-\varphi^*)$, $\p_\mu\varphi^*\p^\mu\varphi = \frac{1}{2} (\p_\mu\Sigma\p^\mu\Sigma + \p_\mu\Pi\p^\mu\Pi)$, and $A^2 = |\varphi|^2$. We may equivalently replace the effective action for $A$ with the one for $\varphi$. The effective Lagrangian for $\varphi$ reads 
 $
    \mathcal{L}_\varphi = \partial_\mu\varphi^*\partial^\mu\varphi - V(|\varphi|^2)\,,
    $
where $V(|\varphi|^2)$ is provided by Eq.~\eqref{effpotential}, $A$ being replaced by $|\varphi|$. The field $\varphi$ has a vector $U(1)$ symmetry, and can be written as 
$
    \varphi(t,x) = \rho(t,x)e^{\imath\theta(t,x)}\,.
    $
Its general features of instability can be studied according to Floquet analysis, following Refs.~\cite{Kusenko:1997si,Cotner:2019ykd}, resulting in a condition of Q-ball (a kind of non-topological solitons) formation, i.e.  
$ V''(\rho) - \dot{\theta}^2 <0\,.$
Thus, when $V''(\rho)<0$ holds, there always exist unstable perturbations that grow to form Q-balls.\\

\begin{figure}[htbp]
  \centering
  \includegraphics[width=0.45\textwidth]{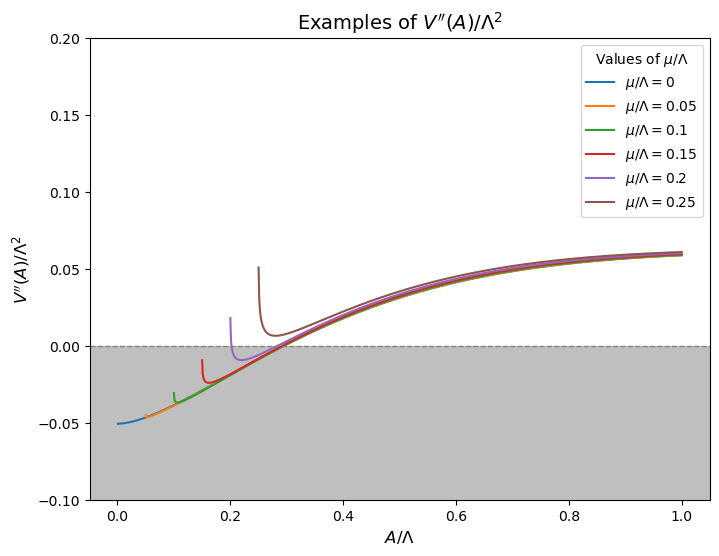}
  \caption{We plot the second derivative of the normalized effective potential for different values of the chemical potential. Regions where the derivative is negative are highlighted in gray, below the dashed horizontal line. When the value of the chemical potential is small, the potential exhibits a region of instability before becoming positive. For larger chemical potential values, the second derivative remains positive throughout, indicating stability. This suggests that at low chemical potential, the effective potential is unstable for small gap field values, but the instability diminishes as the gap field grows.}
  \label{fig:2nddV}
\end{figure}

To investigate Q-balls from the fermion condensate, we inspect the second derivative of the effective potential in Eq.~\eqref{effpotential}, which is plotted in Fig.~\eqref{fig:2nddV}. We then assume that, at the beginning, the axial number density $\langle n_A\rangle \approx 0$, which implies the chemical potential $\mu\approx 0$, supported by inflation, i.e. the number density is diluted during inflation. Along the relaxation of the gap field $A$ toward the minimum of the potential --- the waterfall regime --- we find that the second derivative of the potential changes from negative to positive values when $\mu$ is small. Initially, the instabilities of the field $\varphi$ undergo a dramatic growth, becoming stable when the second derivative becomes positive. This condition triggers the formation of a stable fragmentation, with Q-balls arising from the potential. 

The Q-ball profile can be calculated by minimizing the energy, following e.g. Refs.~\cite{Kusenko:1997zq,Dvali:1997qv}, in which 
\begin{align}\label{QballE}
    E_\omega =&\int d^3x\biggl(\frac{1}{2}\left[\dot{\rho}^2+\rho^2(\dot{\theta}-\omega)^2+(\nabla\rho)^2+\rho^2(\nabla\theta)^2\right]\notag\\
    &+ V(\rho)-\frac{1}{2}\omega^2\rho^2\biggr) +  \omega Q\,, 
\end{align}
where $\omega \equiv \dot{\theta}$ is a Lagrangian multiplier and $Q = \omega \int_{V_3} d^3x \rho^2(x)$.
We then find 
\begin{align}\label{QW}
    \omega \propto \Lambda Q^{-1/4}, \quad  E_Q \propto \Lambda Q^{3/4}, \quad R \propto \frac{Q^{1/4}}{\Lambda},
\end{align}
which aligns with the supersymmetric Q-balls' profile \cite{Kasuya:2000wx,Kasuya:2001hg,Cotner:2019ykd}. Q-balls are not always absolutely stable. They may either decay into lighter fermions \cite{Cohen:1986ct,Kawasaki:2012gk,Enqvist:1998xd}, or lead to baryon number violation \cite{Kusenko:2005du,Kasuya:2014ofa,Cotner:2016dhw,kawasaki2005q}. The Q-balls that form will behave as a matter component in the universe, leading to a brief period of matter domination. As the universe expands and the Q-balls evaporate, radiation once again becomes the dominant component. A detailed discussion and quantitative analysis of this process can be found in Ref.~\cite{Cotner:2019ykd}. Under certain conditions, a fraction of the surviving Q-balls can evolve into primordial structures such as primordial black holes (PBHs).\\
%

\emph{Discussion}. --- 
We have introduced an inflationary model that mirrors key aspects of hybrid inflation, arising from two classes of fermionic species. In this framework, both the inflaton and the auxiliary field responsible for the waterfall transition emerge as composite fields formed from fermion condensates, rather than as ad hoc fundamental scalars. As we have shown, fermion-condensate-driven inflation naturally connects to observable phenomena, including the production of gravitational waves and the abundance of primordial black holes (PBHs) as dark matter. Because inflation here is sourced by self-interacting fermions, its cosmological signatures are intrinsically tied to the physical consequences of these interactions. For instance, gravitational wave damping due to fermion self-interactions and decoupling effects has been explored in~\cite{Loverde:2022wih}. Furthermore, these interactions imply the presence of dynamical Chern-Simons (dCS) gravity \cite{Alexander:2022cow}, where the decay constant relates to the effective mass of self-interacting fermions, establishing a connection between these latter and the dCS framework. Additionally, dCS gravity induces birefringence in primordial gravitational waves, significantly affecting their power spectra in the presence of damping \cite{Liu:2025ifb}. This reinforces the observational relevance of the model. 

\vspace{0.5cm}
\begin{acknowledgments}
J.L. and A.M. wish to thank Cristiano Germani, Tucker Manton, Alexander Vikman and Yingli Zhang for valuable discussions.
The work of S.A.~is supported in part by the Simons Foundation award number 896696.
P.C.~acknowledges the supports by Leung Center for Cosmology and Particle Astrophysics (LeCosPA), National Taiwan University and National Science and Technology Council in Taiwan.
J.L.~is supported by the Natural Science Foundation of China, grant No.BZ4260016.
A.M.~wishes to acknowledge support by the Shanghai Municipality, through the grant No.~KBH1512299, by Fudan University, through the grant No.~JJH1512105, the Natural Science Foundation of China, through the grant No.~11875113, and by the Department of Physics at Fudan University, through the grant No.~IDH1512092/001.
This work is supported in part by JSPS KAKENHI grant No.~JP24K00624, and by the program YITP 25-100.
\end{acknowledgments}

\bibliographystyle{apsrev4-2}
\bibliography{reference}

\end{document}